\documentclass[prl,twocolumn,epsfig,rotate,showpacs]{revtex4}
\usepackage{epsfig}
\usepackage{amsmath}
\usepackage{graphicx}
\usepackage{dcolumn}
\usepackage{bm}

\begin{document}

\title{Energy Diffusion in Gases}
\author{Jinghua Yang }
\author{Yong Zhang}
\author{Jiao Wang}
\author{Hong Zhao}
\email{zhaoh@xmu.edu.cn} \affiliation{
Department of Physics, and Institute of Theoretical Physics and Astrophysics,\\
Xiamen University, Xiamen 361005, China. }
\date{\today}

\begin{abstract}
In the air surrounding us, how does a particle diffuse? Thanks to
Einstein and other pioneers, it has been well known that generally
the particle will undergo the Brownian motion, and in the last
century this insight has been corroborated by numerous experiments
and applications. Another fundamental question is how the energy
carried by a particle diffuses. The conventional transport
theories assumed the Brownian motion as the underlying energy
transporting mechanism, but however, it should be noticed that in
fact this assumption has never been tested and verified directly
in experiments. Here we show that in clear contrast to the
prediction based on the Brownian motion, in equilibrium gases the
energy diffuses ballistically instead, spreading in a way
analogous to a tsunami wave. This finding suggests a conceptually
new perspective for revisiting the existing energy transport
theories of gases, and provides a chance to solve some important
application problems having challenged these theories for decades.
\end{abstract}

 \maketitle

Diffusive motion is fundamental in nature. The conventional
diffusion theory for a particle immersed in a fluid evolves on the
basis of the Einstein's 1905 work on the Brownian
motion\cite{Einstein}. It is realized that the motion of the
particle can be essentially modeled with the random
walk\cite{smoluchowski}, and the corresponding probability
distribution function (PDF) is governed by the diffusion
equation\cite{Einstein,smoluchowski}. This theory predicts the
Gaussian PDF, which represents a general class of slower motions in
nature. In the last century, it has been verified in a wide range of
contexts.  For example, the experiments have shown that under normal
conditions, i.e., under atmospheric condition and at room
temperature, the diffusions of particles are indeed of Gaussian PDF
in a variety of systems such as gases, liquids, surfaces of solids,
and so on. In fact, the diffusion coefficients of gases provided in
manuals of physical properties\cite{experience}, which are useful in
the studies of physics, chemistry, meteorology and other sciences,
are measured based on the conventional diffusion theory.  Due to its
great success, such a particle diffusion picture given by the
conventional diffusion theory has been the common knowledge in
various scientific disciplines.

In a further development of the early transport theory, the clear
physical picture of particle diffusion outlined in Einstein's work
was extended by Helfand\cite{Helfand} in 1960 to link -- through a
set of generalized Einstein relations -- the macroscopic transport
coefficients such as those of molecule diffusion, viscosity and
thermal transportation on one side, to the corresponding
microscopic fluctuating quantities, known as Helfand moments, on
the other side, by expressing the former as the linear time
increase rates of the statistical variances of the latter. Like
the Green-Kubo formula\cite{Green-1951,Green-1960,Kubo}, Helfand's
theory interprets the transport coefficients in terms of the
microscopic dynamics but in a different way, hence provides
additional information and insights. In recent years, besides the
equivalence between the Green-Kubo formula and the Helfand
relations in certain cases, some advantages unique to the Helfand
approach have also been realized\cite{Viscardy-1,Viscardy-2}.

The Helfand theory addressed the transports of the most important
physical quantities including the energy. It is meanwhile the only
theory that suggests an answer to the question we focus on in this
work, i.e., how the energy initially carried by a particle diffuses.
According to the Helfand theory, the energy will diffuse in a random
way and the corresponding PDF is Gaussian as well. This could be
agreed by most scientists nowadays in view of the overwhelming
prevalence of the conventional diffusion theory. However, it should
be pointed out that if this is true has never been studied
experimentally. Unlike in the case of particle diffusion, where the
motion of a particle can be traced accurately, and this is possible
even for the diffusion of a molecule identical to other gas
molecules with the help of labeled atoms, a key difficulty in the
study of energy diffusion is that the energy itself cannot be traced
at all as it will be transferred from molecule to molecule and
shared by more and more molecules via interactions. This difficulty
cannot be overcome even in numerical studies. This is why there does
not exist any reported experimental or numerical data in literatures
for showing how the energy carried by a particle may transport in
gases.

Undoubtedly, the direct evidence is essential to the foundations of
the Helfand energy diffusion theory. In this respect the following
three points deserve particularly careful considerations. First, the
aim of the particle diffusion theory is to address the stochastic
nature of the motion of a single particle, but by nature, the energy
transport should be more complicated because it is essentially a
collective behavior that may involve many gas molecules at a time.
The question, whether -- and if yes to what extent -- a diffusion
theory aiming at addressing a single particle can be extended to
that aiming at addressing a collective behavior of multiple
molecules, has not been answered. Indeed we have good reasons to be
careful as far as this problem is concerned. An illuminating example
is the superfluidity of helium\cite{Landau}, from which we have
learned that the statistical theory based on independent particles
may fail to explain the phenomena whose essence is a collective
behavior. Second, the Helfand theory assumes that the macroscopic
transport behavior, characterized by the linear dependence on the
energy distribution gradient of the energy flux, is still valid on
the microscopic scale. This assumption is not verified, either.
Finally, it should be pointed out that one also must be careful when
applying the Helfand theory to real systems, because to what extent
the fluctuation of a Helfand moment can be characterized by the
random walk thus a linear time increasing variance, is not known
yet. Even for the particle diffusion of a gas molecule, due to the
long time power law velocity autocorrelation\cite{Alder-1970}
observed in the molecular dynamics study, the fluctuation of the
corresponding Helfand moment is not of the rigorous random walk. The
situations for other transports, e.g., the energy as being focused
in this work, should be more complicated because of its collective
behavior nature, which may thus induce both strong time and space
correlations.

We perform an equilibrium molecular dynamics investigation to
snapshoot the process of the energy diffusion directly.  In the
following we will restrict ourselves to a 2D gas model, but it has
been verified that in its 3D counterpart the results reported here
remain to be qualitatively the same. We assume that the gas consists
of only one kind of molecules, and $\sigma$ and $m$ represent the
diameter and the mass of a molecule, respectively. The setup
consists of a square space of area $S$ with periodic boundary
conditions and $N$ molecules moving inside. The interaction between
any two molecules is given by the Lennard-Jones potential and the
Hamiltonian of the system reads
\begin{equation} \label{eq:one}
H=\sum_{i}H_{i}=\sum_{i}\{\frac{\mathbf{p}_{i}^2}{2m}+\sum_{i\neq
j}{2\varepsilon
[(\frac{\sigma}{r_{ij}})^{12}-(\frac{\sigma}{r_{ij}})^{6}]}\},
\end{equation}
where $\varepsilon$ is a constant governing the interaction strength
between molecules and $r_{ij}$ denotes the distance between molecule
$i$ and $j$. Given these the evolution of the system can be
simulated directly. In our calculations the dimensionless parameters
$\varepsilon=1$, $\sigma=1$, $m=1$ and the Boltzmann constant
$k_{B}=1$ are adopted, and the gas density is set to be
$N/S=0.0625$. Another important parameter is the temperature, which
is fixed at $T=2.5$, a value that corresponds to the room
temperature with other adopted dimensionless parameters. To make the
simulations more efficient, the potential energy between two
molecules is approximated by zero when the distance between them is
larger than $r_{c}=3.5$, as conventionally adopted in the molecular
dynamics studies of gases.

We first investigate the particle diffusion. First of all an
equilibrium state of $N=2500$ molecules at temperature $T=2.5$ is
prepared by evolving the system for a long enough time ($>1\times
10^{4}$) from a random initial condition where the positions of the
molecules are randomly and uniformly assigned, and their velocities
are generated from the Maxwell-Boltzmann distribution. Then a
molecule, hereafter named the "tagged molecule", is picked up at
random, and its position is set as the coordinate origin. The
diffusion of the tagged molecule can then be studied by tracking its
ensuing motion. Fig. 1 (a)-(b) presents the PDF of the tagged
molecule, $\rho_{m}(\mathbf{r},t)$, evaluated over an ensemble of
$3\times10^{8}$ independent systems (random realizations) prepared
in the same way. (Note that our ensemble is equivalent to the
"subensemble" as considered by Helfand\cite{Helfand}.) Initially
$\rho_{m}(\mathbf{r},t=0)=\delta(\mathbf{r})$; (see Fig. 1(a)), and
later it evolves into a Gaussian distribution (see Fig. 1(b) for
$\rho_{m}(\mathbf{r},t=15)$; Fig. 1(c) and Fig. 1(d) are the
corresponding contours and the intersection with $y=0$ at $t=15$;
$\mathbf{r}\equiv(x,y)$.) As a double check we have also studied the
behavior of the squared displacement of the tagged molecule and
found it depends on time linearly; i.e.,
$\langle|\mathbf{r}|^{2}(t)\rangle\sim t$. These results are clear
evidence that the molecule diffusion is normal in our system.

\begin{figure*}
\vspace{-.2cm}\hspace{0.cm}\epsfig{file=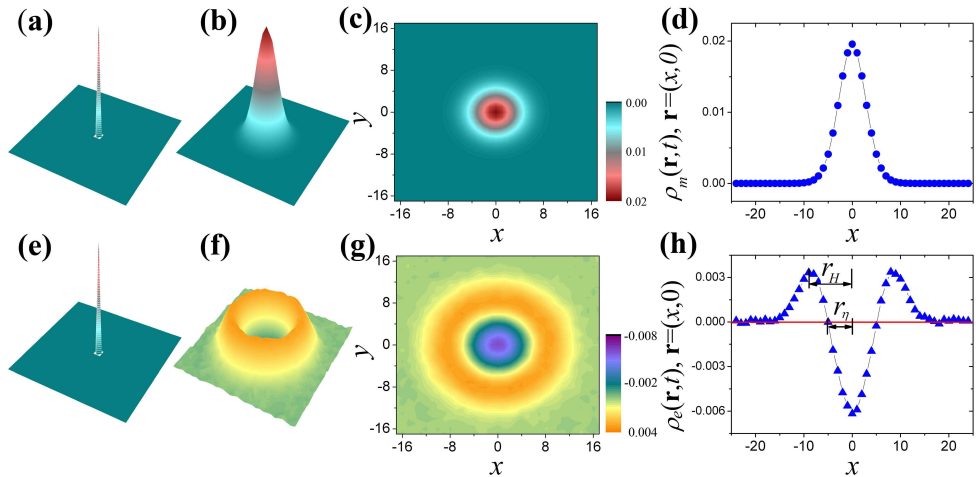,width=13cm}
\caption{The diffusion of a 2D gas molecule measured by the density
distribution $\rho_{m}(\mathbf{r},t)$(a)-(d) and that of the energy
it carries measured by $\rho_{e}(\mathbf{r},t)$(e)-(h). Initially
the molecule is located at the origin (a); $\rho_{m}(\mathbf{r},t)$
at $t=15$ is shown in (b) and its contours and intersection with
$\mathbf{r}=(x,y=0)$ are shown in (c) and (d) respectively. (e)-(h)
are the counterparts of (a)-(d) but for $\rho_{e}(\mathbf{r},t)$
instead. }\label{fig1}
\end{figure*}

However, surprisingly, our next study suggests the energy diffusive
behavior can be significantly different. To study the energy
diffusion a technical difficulty we encounter is that the energy
cannot be tagged like a particle, as it will be shared and
transferred among the molecules with an increasing number due to
interactions. However, in this case the idea of Helfand's
"subensemble", i.e., the ensemble we have previously adopted by
randomly selecting a molecule and setting its position as the
origin\cite{Helfand}, is still valid and useful. We consider the
energy distribution of the system
$E(\mathbf{r},t)=H_{1}\delta(\mathbf{r}-\mathbf{r}_{1})+\sum_{i=2}^{N}H_{i}\delta(\mathbf{r}-\mathbf{r}_{i})$,
where $\mathbf{r}_{i}$is the position of molecule $i$ at time $t$.
In particular, we refer to the first molecule as the tagged molecule
that resides on the origin initially, and focus in the following on
how the total energy it bears (at time $t=0$) diffuses. For this
purpose we take the ensemble average
\begin{equation} \label{eq:two}
\langle E(\mathbf{r},t)\rangle=\langle
H_{1}\delta(\mathbf{r}-\mathbf{r}_{1})\rangle+\langle
\sum_{i=2}^{N}H_{i}\delta(\mathbf{r}-\mathbf{r}_{i})\rangle,
\end{equation}
which gives the energy density distribution of the system.
Initially, as the origin is occupied exclusively by the tagged
molecule, we have $\langle
H_{1}\delta(\mathbf{r}-\mathbf{r}_{1})\rangle=\widetilde{E}\delta(\mathbf{r})$;
here $\widetilde{E}=\langle H_{j}\rangle$ is the average energy of a
molecule in the gas. For the rest area of the space, i.e.,
$\mathbf{r}\neq0$, as it is occupied by the other $N-1$ untagged
molecules uniformly, the energy density distribution they contribute
to, represented by the term $\langle
\sum_{i=2}^{N}H_{i}\delta(\mathbf{r}-\mathbf{r}_{i})\rangle$, equals
a constant $\eta\equiv\widetilde{E}(N-1)/S$. Hence initially
$\langle E(\mathbf{r},t=0)\rangle$ is characterized by a center of
$\delta$-function form and a flat background. This has been well
verified by the simulations (see Fig. 1(e)).  As the system evolves,
while the energy density distribution of the tagged molecule may
spread out from its initial $\delta$-function, that of the other
$N-1$ molecules is expected to be the same as $\eta$ as well. This
leads to the conclusion that a reformed distribution,
$\rho_{e}(\mathbf{r},t)\equiv(\langle
E(\mathbf{r},t)\rangle-\eta)/\widetilde{E}$, can well capture the
diffusion process of the energy the tagged molecule carries
initially. This is the key point of our argument; given it the
measuring of the energy diffusion of a single molecule is reduced to
that of the energy density distribution of the whole system, making
it possible to study the former conveniently.

Fig. 1(f) shows the energy diffusion results given by
$\rho_{e}(\mathbf{r},t)$ at time $t=15$, where a distinctive
difference from the molecule diffusion result of
$\rho_{m}(\mathbf{r},t)$(Fig. 1(b)) can be identified. Instead of a
Gaussian distribution, the function $\rho_{e}(\mathbf{r},t)$ is
characterized by a growing "crater", i.e., a ring ridge (where
$\rho_{e}(\mathbf{r},t)>0$) moves outwards leaving behind a dip
(where $\rho_{e}(\mathbf{r},t)<0$) in center. (See Fig. 1(g)-(h) for
the corresponding contours and the intersection with $y=0$). Two key
geometric parameters of the "crater", denoted by $r_{\eta}$ and
$r_{H}$ respectively, are the radius of the intersection ring on
which $\rho_{e}(\mathbf{r},t)=0$ and that of the top ring of the
ridge where $\rho_{e}(\mathbf{r},t)$ takes the maximum value (see
Fig. 1(h)). We find that they depend on time linearly (see Fig.
2(a)).  It should be noted that however the velocity of the ridge,
represented by $\nu_{H}\equiv dr_{H}/dt$, is different from that of
the opening of the dip, i.e., $\nu_{\eta}\equiv dr_{\eta}/dt$: While
the former is $\nu_{H}\approx1.1\nu_{s}$, the latter is
$\nu_{\eta}\approx0.7\nu_{s}$. Here
$\nu_{s}=\sqrt{c_{p}k_{B}T}/\sqrt{c_{v}m}$ is the sound velocity in
our gas model. On the other hand, as $\langle
E(\mathbf{r},t)\rangle-\eta$ describes how the initial energy
carried by the tagged molecule diffuses, the fact that
$\rho_{e}(\mathbf{r},t)$ has a negative center suggests that,
interestingly, during its diffusion some energy of the neighboring
molecules is brought away in addition. This additional portion of
energy is given by $-E_{-}$, where $E_{-}\equiv
\widetilde{E}\int_{|\mathbf{r}|<r_{\eta}}\rho_{e}(\mathbf{r},t)d\mathbf{r}$.
Similarly, the total positive energy carried by the bulk of the
ridge is given by $E_{+}\equiv
\widetilde{E}\int_{|\mathbf{r}|>r_{\eta}}\rho_{e}(\mathbf{r},t)d\mathbf{r}$.
Due to the conversation of the energy, we have always
$E_{+}+E_{-}=\widetilde{E}$. Fig2. (b) shows the time dependence of
$E_{-}$ and that of $E_{+}$; initially $E_{-}$($E_{+}$) decreases
(increases) but after a transition time it approach a constant. In
other words, eventually the total energy brought away by the ridge
is a constant and larger than the energy initially the tagged
molecule carries. Together with the results of $\nu_{H}$ and
$\nu_{\eta}$, they suggest clearly that rather than Gaussian, the
energy diffusion follows a ballistic way resembling the process that
the tsunami waves transport away the energy released from a sea
earthquake.

Now let us explain why the two diffusion behaviors are so different.
Consider a Brownian particle, e.g., our tagged molecule; Due to its
frequent collisions with other molecules, its memory of the initial
direction of motion suffers a quick loss. This process can be
measured by the decay of the autocorrelation function
$A(t)\equiv\langle\mathbf{p}_{1}(0)\cdot\mathbf{p}_{1}(t)\rangle$ of
the tagged molecule. Here $\mathbf{p}_{i}(t)$ is the momentum of the
molecule $i$ and $\mathbf{p}_{1}(t)$ is that of the tagged molecule.
Indeed, the simulation suggests that $A(t)$ decreases exponentially
in time in our model (not shown here), hence the motion of the
tagged molecule is essentially equivalent to that of a random
walker. This explains why the normal particle diffusion is observed.
However, as the momentum of the tagged molecule can be transferred
to other molecules during the interactions, the memory of it
"remembered" by all the molecules should thus be measured by the
correlation function
$M(t)\equiv\sum_{j}\langle\mathbf{p}_{1}(0)\cdot\mathbf{p}_{j}(t)\rangle$
evaluated over the whole system. Dividing the momentum of a
molecule, say molecule $j$, into two parts:
$\mathbf{p}_{j}(t)=\mathbf{p}_{j}'(t)+\mathbf{p}_{j}''(t)$, where
$\mathbf{p}_{j}'(t)$($\mathbf{p}_{j}''(t)$) represents the momentum
transferred to it from the tagged molecule (other molecules), we
then have
$M(t)=\sum_{j}\langle\mathbf{p}_{1}(0)\cdot\mathbf{p}_{j}'(t)\rangle=\langle\mathbf{p}_{1}(0)\cdot\mathbf{p}_{1}(0)\rangle$;
This is because (1)
$\langle\mathbf{p}_{1}(0)\cdot\mathbf{p}_{j}''(t)\rangle=0$ since
$\mathbf{p}_{j}''(t)$ is independent of $\mathbf{p}_{1}(0)$ and (2)
$\sum_{j}\mathbf{p}_{j}'(t)=\mathbf{p}_{1}(0)$ since the momentum
$\mathbf{p}_{1}(0)$ is conserved in the system. This result that
$M(t)$ is in fact a time-independent constant suggests that, though
the information of its initial state will be forgotten quickly by a
molecule itself, it will always be remembered by others. Hence it is
in effect not lost.

\begin{figure}
\epsfig{file=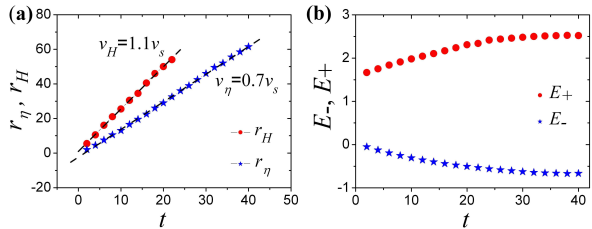,width=8cm} \caption{(a) The time dependence of
the characterizing radii, corresponding to the top ring of the ridge
(red bullets) and the opening of the center dip (black stars),
respectively. The best fittings (dashed lines) suggest their
expending speeds $\nu_{H}\approx1.1\nu_{s}$ and
$\nu_{\eta}\approx0.7\nu_{s}$. (B) The positive and negative potion
of energy, $E_{+}$(red bullets) and $E_{-}$(black stars),
corresponding to the integrated energy density distribution over
region $|\mathbf{r}|>r_{\eta}$ and $|\mathbf{r}|<r_{\eta}$
respectively. }\label{fig2}
\end{figure}

In Fig. 3 presents the simulation results of
$\rho_{m}(\mathbf{r},t)$ and $\rho_{e}(\mathbf{r},t)$ for an
ensemble where the momentum direction of the tagged molecule is
normalized as well. Unlike in the study presented in Fig. 1, where
for each random realization the tagged molecule has a momentum with
random direction and hence the memory effects of the direction are
hidden, here we instead reset (renormalize) the direction of the
initial momentum of the tagged molecule to be the same as the axis
$y$. This is equivalent to considering a subset of the Helfand
subensemble where the direction of the momentum of the considered
molecule is specified as well. As such the tagged molecule will keep
going along axis $y$ in a short initial stage. It can be seen from
Fig. 3 that while $\rho_{m}(\mathbf{r},t)$ is isotropic, implying a
memory loss effect, $\rho_{e}(\mathbf{r},t)$ is obviously
anisotropic, showing a strong signal of the initial moving direction
of the tagged molecule.

Because the memory is kept during the momentum diffusion and thus
the energy diffusion process (as the energy is transferred
simultaneously with the momentum), the energy diffusion cannot be a
Markov process. This explains why it shows an abnormal diffusion
behavior. However, at present we cannot explain yet why the
diffusion is ballistic, which calls for further studies in future.
It should be noted that in the macroscopic world the ballistic
transport of energy has been found ubiquitous. For example, it is in
a ballistic way that the shock waves bring away the energy of an
explosion, and so do the sea waves in a tsunami to bring away the
energy from a sea earthquake. Even in a much more "peaceful" case
like dropping a pebble into a pond, it is the way that partial
kinetic energy of the pebble is carried away by the homocentric
ripples. In all these examples, it is certain that the bulk of the
excited energy is transported by waves. Our finding in this Letter
implies that the wave could also be a general energy transporting
approach on the microscopic level, i.e., on a single molecule level
as exposed here. The only differences of our finding from the
macroscopic examples cited above lies in that our results are based
on the ensemble average.

\begin{figure}
\epsfig{file=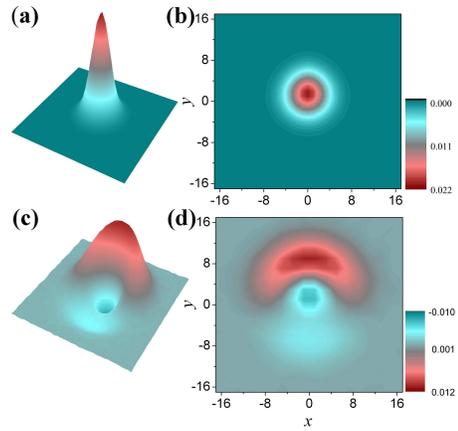,width=6cm} \caption{The density distributions
of a 2D gas particle $\rho_{m}(\mathbf{r},t)$(a)-(b) and the related
energy density distribution $\rho_{e}(\mathbf{r},t) t=15$(c)-(d).
Initially the particle is located at origin, and its velocity
direction is renormalized to be the same as axis $y$. The right
column is the contours plot of the corresponding figure in left
column. $3\times10^{8}$random realizations are considered for
ensemble average. }\label{fig3}
\end{figure}

In summary, we have shown numerically that the energy carried by a
molecule in a 2D gas may spread out in a ballistic way at room
temperature. Considering the ensemble average, the energy profile is
found to be characterized by a ring ridge and a dip in center, and
both expand outward with constant speeds. As a basic mechanism of
the energy transportation in gas, we believe this property may find
important applications. For example, one possible situation is the
Tokamak plasma, where the heat conductivity has been found to
deviate significantly from the prediction based on classical normal
diffusion theories\cite{Chen-f-f}. Our finding provides a new
perspective for revisiting such challenging problems.

This work is supported by the National Natural Science Foundation of
China under Grant No. 10775115, 10925525, and the National Basic
Research Program of China (973 Program) (2007CB814800).

\end{document}